\documentclass[reqno,11pt]{article}
\usepackage{amssymb,amsthm,amsmath}
\usepackage{epsfig}

\theoremstyle{plain}
\begingroup
 
\newtheorem{thm}{Theorem}[section] 

\newtheorem{lemma}[thm]{Lemma}

\endgroup

\newcommand{\nwc}{\newcommand}
\nwc{\qref}[1]{(\ref{#1})}
\nwc{\z}{z}
\nwc{\D}{\partial}
\nwc{\C}{{\mathbb{C}}}
\nwc{\Cint}{\mathbb{C}_+}
\nwc{\Cclose}{\bar{\mathbb{C}}_+}
\nwc{\R}{{\mathbb{R}}}
\nwc{\Z}{{\mathbb{Z}}}
\nwc{\N}{{\mathbb{N}}}
\nwc{\LK}{L\'{e}vy-Khintchine}
\nwc{\levy}{L\'evy}
\nwc{\Gammamt}{\Gamma_{-t}}
\nwc{\Phimt}{\Phi^{-1}_t}
\renewcommand{\Re}{\mathop{\rm Re}\nolimits}
\renewcommand{\Im}{\mathop{\rm Im}\nolimits}
\nwc{\nsg}[1]{\hat{n}_{*,#1}}
\nwc{\ta}{t} 
\nwc{\tb}{\tau} 
\nwc{\vp}{\varphi}
\nwc{\veps}{\varepsilon}
\nwc{\taus}{\tau_0} 
\nwc{\be}{\begin{equation}}
\nwc{\ee}{\end{equation}}
\nwc{\ba}{\begin{eqnarray}}
\nwc{\ea}{\end{eqnarray}}
\nwc{\nn}{\nonumber}
\nwc{\hx}{\hat{x}}
\nwc{\hn}{\hat{n}}
\nwc{\la}{\label}
\nwc{\etalap}{N}
\nwc{\etaadd}{\Phi}
\nwc{\etamult}{\Psi}
\nwc{\Tgel}{T_{\rm gel}}
\nwc{\att}{a}
\theoremstyle{definition}

\theoremstyle{remark}
 
\numberwithin{equation}{section}
\numberwithin{figure}{section}

\begin{document}
\title{Dynamical scaling in
  Smoluchowski's coagulation equations: uniform convergence} 
\author{Govind Menon\textsuperscript{1} and Robert. L.
Pego\textsuperscript{2}}

\date{\today}
\maketitle

\begin{abstract}
We consider the approach to self-similarity (or dynamical scaling) in
Smoluchowski's coagulation equations  
for the solvable kernels $K(x,y)=2,x+y$ and $xy$.
We prove the uniform convergence of densities to the self-similar
solution with exponential tails under the regularity hypothesis  that
a suitable moment have an integrable Fourier transform. For the
discrete equations we prove uniform convergence under optimal moment
hypotheses. Our results are completely analogous to classical 
local convergence theorems for the normal law in 
probability theory. The proofs rely on the Fourier inversion formula and the
solution by the method of characteristics for the Laplace transform.
\end{abstract}

\footnotetext[1]
{Department of Mathematics, University of Wisconsin, Madison WI 53706.
{\em Current address:\/}
Division of Applied Mathematics, Brown University, Providence, RI 02912.
Email: menon@dam.brown.edu}
\footnotetext[2]{Department of Mathematics \&
Institute for Physical Science and Technology,
University of Maryland, College Park MD 20742. 
Email: rlp@math.umd.edu}
 
\section{Introduction}
Smoluchowski's coagulation equation
\begin{equation}
\label{eq:smol1}
{\partial_t n}(t,x) = \frac{1}{2} \int_0^x K(x-y,y) n(t,x-y)
n(t,y) dy  - \int_0^\infty K(x,y) n(t,x) n(t,y) dy,
\end{equation}
is a widely studied mean-field model for cluster growth~\cite{Drake,
  Friedlander,Smol}. We study the evolution of 
$n(t,x)$, the number of clusters of mass $x$  per
unit volume at time $t$, which coalesce by binary collisions with a symmetric
rate kernel $K(x,y)$. Equation~(\ref{eq:smol1}) has been used as a model
 of cluster growth in a surprisingly 
 diverse range of fields such as  physical chemistry, astrophysics,
  and population dynamics (see~\cite{Drake} for a review of
  applications). In addition, over the past few years a rich
  mathematical theory has been developed for these equations.
 Aldous \cite{Aldous} provides an excellent introduction.

Many kernels in applications are
homogeneous, that is  $K(\alpha x, \alpha y) = \alpha^\gamma K(x,y)$,
  $x,y,\alpha >0$,
for some exponent $\gamma$~\cite{Drake}. A mathematical 
problem of scientific interest is to study self-similar or
dynamical scaling behavior for homogeneous kernels. There are no general
mathematical results for this problem despite an extensive scientific
literature (especially formal asymptotics and
numerics~\cite{Lee,Leyvraz,DE1}). It is known that $\gamma$ plays a crucial
role.  On physical grounds, we expect solutions to (\ref{eq:smol1}) to
conserve the total mass $\int_0^\infty xn(t,x) dx$. When
$K(x,y)\le 1+x+y$ (corresponding to $0\le\gamma \leq 1$),
mass-conserving solutions exist globally in time under suitable moment
hypotheses on initial data~\cite{Dub1}. 
It is then typical in applications to assert
that the solutions approach ``scaling form''~\cite{Leyvraz,DE1}, but there 
is no rigorous mathematical justification for this in general.  

For a large class of kernels satisfying 
$(xy)^{\gamma/2}\le K(x,y)$ with $1<\gamma<2$, 
it is known that there is
no solution that preserves mass for all time. 
This breakdown phenomenon is known as gelation. 
It was first demonstrated by McLeod~\cite{McL} with an explicit solution
for the kernel $K=xy$. A general result using only the
growth of the kernel was proved probabilistically by Jeon~\cite{Jeon}
(see also~\cite{Escobedo} for a simple analytical proof). It is
natural to ask whether the blow-up is self-similar, but there are no
general results on this problem yet.  

There are a number of results, however, for the `solvable' kernels
$K=2$, $x+y$ and $xy$ 
(see~\cite{MP1} and references therein; also see \cite{Leyvraz}).
A remarkable feature of these kernels is that the problem of dynamical
scaling can be understood quite deeply by analogy with classical
limit theorems in probability theory.
For example, an analog to the classical \LK\/ representation for infinitely
divisible laws was proved by Bertoin~\cite{Bertoin} 
for eternal solutions to Smoluchowski's equation with kernel $K=x+y$. 
Eternal solutions are defined for all
$t\in(-\infty,\infty)$, meaning they model coagulation processes
`infinitely divisible' under Smoluchowski dynamics.
Later, we proved~\cite{MP1} that the domains of attraction of 
self-similar solutions (in the sense of weak convergence of measures) 
can be characterized by almost power-law behavior of the tails of the 
initial size distribution. This is
analogous to the characterization of the weak domains of
attraction of the \levy\ stable laws~\cite{Feller}. 
An essential component in both proofs is a simple solution formula
for the Laplace transform of $n$ that is widely known~\cite{Drake}.
These results may be used as a basis for refined convergence theorems, 
as we now explain.

A general theme in probabilistic limit theorems is the interplay between
moment  and regularity hypotheses and the topology of
convergence. In this article, we develop one aspect of this
idea. Under stronger regularity hypotheses, the
weak convergence results of~\cite{MP1} will be
strengthened  to obtain uniform convergence of densities using the Fourier
transform. This method is classical in probability
theory and is used to prove uniform 
convergence of densities in the central limit
theorem~\cite[XV.5.2]{Feller}. Feller's argument in \cite{Feller} 
is simple and  robust, and our main contribution is to show that it 
extends naturally to Smoluchowski's equation. 
The key new idea is to use the method of characteristics in the right half
of the complex plane to obtain strong decay estimates on the Laplace 
transform.  
A broader contribution of this work and~\cite{MP1} is to show that
the analytical methods used to prove classical limit theorems in
probability apply to a wider range of problems involving
scaling phenomenon for integral equations of convolution type. 

Let us briefly connect our results to previous work: the only uniform
convergence theorems in the literature are that of Kreer and
Penrose for the kernel $K=2$~\cite{KP}, and closely connected work of
daCosta~\cite{daCosta}. In this article, 
for $K=2$ and $x+y$ we present theorems on uniform convergence 
to the self-similar solutions with exponential
tails for the continuous and discrete Smoluchowski equations. 
For $K=xy$, we prove uniform convergence of densities to self-similar 
form  as $t$ approaches the gelation time $\Tgel$.
For $K=2$, we strengthen the result of Kreer and Penrose and simplify
the proof. Their decay hypothesis on the initial data ($n_0(x) \leq Ce^{-a x}$) 
is weakened to an (almost) optimal moment hypothesis, and
their regularity hypothesis ($n_0 \in C^2$) is weakened to a little bit more
than continuity. For $K=x+y$ the convergence theorem is new. 
Study of the kernel $K=xy$ is reduced to $K=x+y$ by 
a well-known change of variables~\cite{Drake}. 
Uniform convergence to the self-similar solutions with ``fat'' or
``heavy''  tails is a more delicate issue, and will not be considered here.

Our uniform convergence theorems may be stated in a unified manner as follows 
for the continuous Smoluchowski equations with kernels 
$K(x,y)=2$, $x+y$ and $xy$, corresponding to $\gamma=0,1,2$ respectively.
Presuming the $\gamma$-th and $(\gamma+1)$-st moments are finite, 
we may scale $x$ and $n$ so 
both moments are initially 1. For the multiplicative
kernel this ensures that the gelation time $\Tgel=1$.
Let $T_\gamma=\infty$ for $\gamma =0,1$, 
$T_\gamma=T_{\rm gel}=1$ for $\gamma =2$.
The self-similar solutions with exponential tails are explicitly 
given by~\cite{MP1}
\begin{equation}\label{eq:sss}
n(t,x) =
\frac{m_\gamma(t)}{\lambda_\gamma(t)^{\gamma+1}}
\,\nsg{\gamma}\!\left(\frac{x}{\lambda_\gamma(t)}\right),
\end{equation}
where for $\hat{x}\ge0$,
\begin{equation}
\nsg{0}(\hat{x}) = e^{-\hat{x}}, \qquad \hat{x} \nsg1(\hat{x}) = 
\hat{x}^2\nsg2(\hat{x})=
\frac{1}{\sqrt{2\pi}}\hat{x}^{-1/2}e^{-\hat{x}/2}, 
\label{eq:ndef}
\end{equation}
and
\begin{eqnarray}
&& m_0(t) = t^{-1},\quad m_1(t)=1, \quad m_2(t)=(1-t)^{-1},
\label{eq:mdef}\\
&& \lambda_0(t)=t, \quad \lambda_1(t)=e^{2t}, \quad
\lambda_2(t)=(1-t)^{-2}. 
\label{eq:ldef}
\end{eqnarray}
Our sufficient conditions for uniform convergence to these self-similar
solutions for the continuous Smoluchowski equations are summarized by
the following result.
\begin{thm}
\label{thm:main}
Let $n_0 \geq 0$,  $\int_0^\infty x^{\gamma} n_0(x) dx =   
\int_0^\infty x^{1+\gamma} n_0(x) dx =1$. Assume that the Fourier
transform of $x^{1+\gamma}n_0$ is integrable, and let 
$n(t,x)$ be the solution  to Smoluchowski's equation 
with initial data $n_0(x)$ and
$K=2$, $x+y$ or $xy$, for $\gamma=0,1$ or $2$. 
Then the rescaled solution
\[
\hat{n}(t,\hat x) = 
\frac {\lambda_\gamma(t)^{1+\gamma}} {m_\gamma(t)}
{n}(t,\hat{x}\lambda_\gamma(t)) 
\]
satisfies
\[ \lim_{t \to T_\gamma} \,\sup_{\hat{x}>0}\,
\hat{x}^{1+\gamma}
\left|
\hat{n}(t,\hat{x})- \nsg{\gamma}(\hat{x}) \right| =0. \]
\end{thm}

It has been traditional to treat the discrete Smoluchowski equations
separately from the continuous equations. Yet,
within the framework of measure valued solutions~\cite{MP1,Norris}, the 
discrete Smoluchowski equations simply correspond to the special
case of a lattice distribution, a measure valued solution supported on
the lattice $h\mathbb{N}$ and taking the
form
$\nu_t=\sum_{l=1}^\infty n_l(t)\delta_{hl}(x)$, where $\delta_{hl}(x)$
is a Dirac delta at $hl$. If $h$ is maximal we call $\nu_t$ a 
lattice measure with {\em span} $h$.  
The coefficients $n_l$ satisfy the discrete
Smoluchowski equations
\begin{equation}\label{eq:disc-smol}
\partial_t n_l(t) = \frac{1}{2} \sum_{j=1}^{l-1} \kappa_{l-j,j} 
n_{l-j}(t)n_j(t)
 -\sum_{j=1}^\infty \kappa_{l,j}n_l(t)n_j(t),
\end{equation}
where $\kappa_{l,j}=K(lh,jh)$.
Physically, this case is of importance, since 
some mass aggregation processes (e.g., polymerization) have a fundamental unit
of mass (e.g., a monomer).  The uniform convergence theorems for the
continuous Smoluchowski equations have a natural
extension to this case. 
\begin{thm}
\label{thm:main_discrete}
Let $\nu_0 \geq 0$ be a lattice measure with span $h$ such that
$\int_0^\infty x^{\gamma}\nu_0(dx) = \int_0^\infty x^{1+\gamma}
\nu_0(dx)=1$. Then with 
\[ \hat{l}=\frac{lh}{\lambda_\gamma(t)} , \quad
\hat{n}_l(t)=\frac1h \frac{\lambda_\gamma(t)^{1+\gamma}}{m_\gamma(t)}
n_l(t),
\]
we have
\begin{equation}
\nonumber
\lim_{t \rightarrow T_\gamma} \,\sup_{l\in\N}\,
 \hat{l}^{1+\gamma}\left|  \hat{n}_{l}(t)  
- \nsg{\gamma}(\hat{l}) \right| =0.
\end{equation}
\end{thm}
Let us comment on the hypotheses in Theorems~\ref{thm:main} and
\ref{thm:main_discrete}. The moment  hypotheses in both theorems are
essentially the same. $\int_0^\infty x^{\gamma}\nu_0(dx)=1 $ is
the natural hypothesis for existence and uniqueness of 
solutions~\cite{MP1}. The other moment
condition $\int_0^\infty x^{1+\gamma} 
\nu_0(dx)=1$ is of a different character. 
It implies that $n_0$ or $\nu_0$ is in the weak domain of
attraction of the self-similar solution with exponential tail,
under a rescaling
$n(t,x) \longrightarrow \hat{n}(\hat{t},\hat{x})$ that fixes both moments
\[
\int_0^\infty \hat{x}^{\gamma} \hat{n}(\hat{t},\hat{x}) d\hat{x}=
\int_0^\infty \hat{x}^{\gamma+1} \hat{n}(\hat{t},\hat{x})
d\hat{x}=1 
\quad\mbox{for all $\hat{t} \geq 0$.}
\]
The hypothesis that the $(\gamma+1)$-st moment is finite is 
almost optimal. The weak domain of attraction under a broader class of
rescalings is a bit bigger, as it allows for a weak divergence 
$\int_0^y x^{1+\gamma}
\nu_0(dx) \sim L(y)$ as $y \to \infty$ for a slowly varying function
$L(y)$~\cite{MP1}. 
Thus, Theorem~\ref{thm:main_discrete} shows that
within the class of lattice measures, the weak convergence of
measures almost implies uniform convergence of the coefficients.

Theorem~\ref{thm:main} requires an additional hypothesis on
integrability of a suitable Fourier transform. This is a regularity
hypothesis that is the analog of the 
hypothesis for uniform convergence to the normal law used by
Feller~\cite{Feller}. One may heuristically understand the role of
regularity  as follows. Equation (\ref{eq:smol1}) is
hyperbolic and discontinuities in the initial data persist for all finite 
times. On the other hand, the self-similar solutions in
(\ref{eq:ndef}) are analytic.  Thus, 
one expects some regularity on the initial data is necessary to obtain uniform
convergence to a self-similar solution. Loosely speaking, regularity
of the initial data $n_0(x)$ translates into a decay hypothesis on its
Fourier transform. We need only the weak decay implied by
integrability. 

We do not know if this assumption is optimal, or if it may be weakened
further. We briefly comment on 
this issue here; it will not be considered in the rest of the paper.
The space of functions with integrable Fourier transforms is of great
interest in harmonic analysis. Precisely, for $f\in L^1(\mathbb{R})$, let $F$
be its Fourier transform. Then the space 
\[
 A(\mathbb{R}) = \{f \in L^1(\R)|
F \in L^1(\R)\}
\]
is a closed subalgebra of $L^1(\R)$ 
known as the Wiener algebra~\cite{Katznelson}. 
Integrability of $F$ implies
that $f$ is continuous. But it also implies more.
It is known that functions in $A(\mathbb{R})$ possess some
delicate regularity 
properties. For example, a function in $A(\mathbb{R})$ has a
logarithmic modulus 
of continuity in a neighborhood where it is  monotonic. It is
definitely not obvious whether this regularity is truly necessary to obtain
uniform convergence. If $v_0(ik) = \int_0^\infty e^{-ikx}x^{1+\gamma}
n_0(x)dx$ is integrable it also follows that
$v_0 \in H^1(\mathbb{R}) \cap
A(\mathbb{R})$, since $v_0$ is the  boundary limit of an analytic
function (the Laplace transform of $x^{1+\gamma}n_0$). Here $H^1$
denotes the classical Hardy space. This is turn
means that $v_0$ has some hidden regularity and integrability
properties. 
It is worth remarking  that the precise characterization of
$A(\mathbb{R})$ remains an outstanding 
open problem in harmonic analysis (though several sufficient conditions
are known, see~\cite{Katznelson}). 

\section{Uniform convergence of densities for  the constant kernel $K=2$}
\label{sec:const_strong}
\subsection{Evolution of the Laplace transform}
Let $\Cint = \{z \in \C \mid \Re \z >0 \}$ and 
$\Cclose = \{\z \in \mathbb{C}\mid \Re \z \geq 0 \}$. We let
\[
\etalap(\ta,\z) =\int_0^\infty e^{-\z x} n(\ta,x)\, dx, \quad \z \in \Cclose, 
\]
denote the Laplace  transform of the number density $n$. 
We take the Laplace transform of (\ref{eq:smol1}) with $K=2$, and its
limit as $\z \to 0$  
to see that $\etalap(\ta,\z)$ solves 
\begin{equation}
\label{eq:const_1}
\partial_\ta \etalap = \etalap^2 - 2\etalap(\ta,0) \etalap, \quad
\partial_\ta \etalap(\ta,0) = -\etalap(\ta,0)^2 .
\end{equation}
Without loss of generality, we may suppose that the initial time
$\ta=1$. We will always assume that the initial data is normalized such that
\begin{equation}
\label{eq:moment_const}
\int_0^\infty n(1,x) \,dx = \int_0^\infty x n(1,x) \,dx =1. 
\end{equation}
If the initial number of clusters, $\int_0^\infty n(1,x) dx$, and the
mass, $\int_0^\infty x n(1,x) dx$, are finite, we may always assume 
(\ref{eq:moment_const}) holds after rescaling $x$ and $n$.
We solve the second equation in (\ref{eq:const_1}) to see that
the total number of clusters decreases according to 
\begin{equation}
\label{eq:number_const}
 \int_0^\infty n(t,x) \,dx = \etalap(\ta,0)= \ta^{-1}, \quad t \geq 1. 
\end{equation}
We hold $\z$ fixed and integrate (\ref{eq:const_1}) in $t$ to obtain
the solution
\begin{equation}
\label{eq:const_sol}
 \etalap(\ta, \z) = \frac1\ta
\frac{\etalap(1,\z)}{\ta(1-\etalap(1,\z))+\etalap(1,\z)}.
\end{equation}
The evolution preserves mass. Indeed, if we differentiate
(\ref{eq:const_sol}) with respect to $\z$, we find
\begin{equation}
\label{eq:mass_const}
\int_0^\infty x n(t,x) \,dx = -\D_\z \etalap(\ta, 0) = -\D_\z
\etalap(1,0)= \int_0^\infty x n(1,x) \,dx =1. 
\end{equation}

\subsection{Approach to self-similarity}
A special case of the  weak convergence result of~\cite{MP1}, also
given by Leyvraz~\cite{Leyvraz}, is obtained as follows: 
Observe that for each fixed $s \in \Cclose$
equations~(\ref{eq:number_const}), (\ref{eq:const_sol}), and
(\ref{eq:mass_const}) imply  
{\nwc{\etaone}{\etalap(1,s\ta^{-1})}
\begin{equation}
\label{eq:weak1}
\ta \etalap(\ta, {s}{\ta^{-1}}) 
=\frac{\etaone}
{\ta(1-\etaone)+\etaone}
\ \mathop{\longrightarrow}_{\ta\to\infty}\ \frac{1}{1+s}.
\end{equation}
It is classical that the pointwise convergence of Laplace transforms
is equivalent to weak convergence of 
measures~\cite[XIII.1.2a]{Feller}.  Thus,
(\ref{eq:weak1}) implies that rescaled solutions to
Smoluchowski's equations converge weakly. Let us be more precise about
the rescaling. We define the similarity variables 
\begin{equation}
\label{eq:str}
\tb = \log \ta, \quad \hx = \frac{x}{t} = e^{-\tb}x, \quad s = t z
= e^{\tb} z,
\end{equation}
and the rescaled number distribution,
\begin{equation}
\label{eq:rsc}
\hn(\tb,\hx) = e^{2\tb}n(e^\tb,e^\tb\hx) = \ta^2 n(\ta,x). 
\end{equation}
Observe that this rescaling preserves {\em both\/} total number and
mass, that is  
\begin{equation}
\label{eq:mass_preserve}
\int_0^\infty \hn(\tb, \hx) d\hx= \int_0^\infty \hx
\hn(\tb, \hx) d\hx =  1, \quad \tb \geq 0.
\end{equation}
We denote the Laplace transform of $\hn(\tb, \hx)$ by
\begin{equation}
\label{eq:str1}
\quad u(\tb,s) = \int_0^\infty e^{-s\hx} \hn(\tb, \hx)
\,d\hx = e^\tb \etalap(e^\tb, se^{-\tb}) = \ta \etalap(\ta,z).
\end{equation}
In these variables, the pointwise convergence of (\ref{eq:weak1})
takes the simple form
\begin{equation}
\label{eq:weak3} \lim_{\tb \to \infty} u(\tb,s) = \frac{1}{1+s} =:
u_{*,0}(s), \quad 
s \in \Cclose,
\end{equation}
where $u_{*,0}(s)$ denotes the Laplace transform of 
\begin{equation}
\label{eq:weak4}
\nsg{0}(\hx) = e^{-\hx}, \quad \hx \geq 0,
\end{equation}
the profile for the self-similar solution in (\ref{eq:sss}). Now,
(\ref{eq:weak3}) is equivalent to 
\[ \hn(\tb,\hx) \,d\hx \to \nsg{0}(\hx) \,d\hx.\]
as $\tb\to\infty$, in the sense of weak convergence of measures.

Our goal is to strengthen this to uniform convergence in both
continuous and discrete cases, under appropriate hypotheses on initial data. 
For the continuous Smoluwchowski equation \qref{eq:smol1} we prove
\begin{thm}
\label{thm:strong_const}
Let $n(1,x) \geq 0$, 
$ \int_0^\infty n(1,x) \,dx = \int_0^\infty x n(1,x) \,dx =1$. 
Assume  that the Fourier transform  of $xn(1,x)$ 
is integrable.  Then in terms of the rescaling
in (\ref{eq:str})--(\ref{eq:rsc}) we have 
\begin{equation}
\label{eq:strong_fourier}
\lim_{\tb \to \infty} \,\sup_{\hx>0}\,
\hx|\hn(\tb,\hx) - \nsg{0}(\hx) | = 0,
\end{equation}
where $\nsg{0}(\hx) = e^{-\hx}$ 
is the similarity profile in (\ref{eq:weak4}).
\end{thm}
The proof of this theorem extends to treat uniform convergence of
coefficients for solutions of the discrete equations
\qref{eq:disc-smol} under only the hypothesis that the zeroth and
first moments are finite; see Theorem~\ref{thm:const_discrete} below. 

Observe that we prove uniform convergence of the weighted densities 
$\hx \hn(\tb,\hx)$.  
The reason can be ascribed to use of the Fourier-Laplace
inversion formula.  We cannot apply  the inversion
formula directly to $u_{*,0}$ as it is not 
integrable on the imaginary axis ($|u_{*,0}(ik)| \sim
|k|^{-1}$ as $|k|\to\infty$). The slow decay of the Fourier transform
is caused by the jump discontinuity at $x=0$, since $\nsg{0}(x) =0$ for $x
< 0$. In order to gain a uniform convergence result,
we smooth this discontinuity and consider the mass
density $\hx \hn$. Its Laplace transform we denote by
\begin{equation}
\label{eq:def_v_const}
v(\tb, s) = -\D_s u(\tb,s) = 
\int_0^\infty e^{-s\hx} \hx
   \hn(\tb, \hx) \,d\hx. 
\end{equation}
Differentiating (\ref{eq:weak3}), we obtain a corresponding
self-similar profile, with
\begin{equation}
\label{eq:weak2}
v_{*,0}(s) := \frac{1}{(1+s)^2}, \qquad |v_{*,0}(ik)| =
\frac{1}{1+k^2}, \ k \in \mathbb{R}. 
\end{equation}
\subsection{Evolution on characteristics}
The explicit solution for $u(\tb,s)$ and $v(\tb,s)$ can be obtained directly 
by substituting
(\ref{eq:str1}) into (\ref{eq:const_sol}). But we  rederive 
the solution to make explicit the geometric idea underlying the
proof of Theorem~\ref{thm:strong_const}. The same ideas underlie the
proof of Theorem~\ref{thm:strong_add} for the additive kernel
and are more easily understood here. We use the change of
variables (\ref{eq:str}) and 
(\ref{eq:str1}) in (\ref{eq:const_1}), and the conservation of moments
in (\ref{eq:mass_preserve}), to obtain
the equation of evolution for $u$: 
\begin{equation}
\label{eq:str2}
\partial_\tb u + s \partial_s u = -u(1-u).
\end{equation}
The solution of equation (\ref{eq:str2}) may be described by the method of
characteristics. A characteristic curve $s(\tb;\tb_0,s_0)$ is the solution
to
\begin{equation}
\label{eq:str3}
\frac{ds}{d\tb} = s, \quad s(\tb;\tb_0 ,s_0) =s_0 \in \Cclose.
\end{equation}
Explicitly,
\begin{equation}
\label{eq:sol_s}
s(\tb;\tb_0,s_0) = e^{\tb-\tb_0} s_0.
\end{equation}
Equation (\ref{eq:str3}) is an autonomous differential equation in $\Cclose$, 
and may be thought of geometrically. 
For fixed $s_0 \in \Cclose$ the trajectory of the characteristic curve 
$s(\tb;\tb_0,s_0), \tb \in \R$, is a
ray in $\Cclose$ emanating from the origin. In particular, the imaginary axis
is invariant under the flow of (\ref{eq:str3}). Equation
(\ref{eq:sol_s}) shows that the characteristics expand uniformly
outward at
the rate $e^\tau$. Along characteristics we have
\begin{equation}
\label{eq:str4_a}
\frac{du}{d\tb} = -u(1-u), 
\end{equation}
which may be integrated to obtain the solution
\begin{equation}
\label{eq:str4}
u(\tb,s) =
\frac{u(\tb_0,s_0)e^{-(\tb-\tb_0)}}{1-u(\tb_0,s_0)(1-e^{-(\tb-\tb_0)}) }. 
\end{equation}
We need to estimate the decay of the derivative $v = -\partial_s u$. 
Differentiating equation (\ref{eq:str2}), we see that on
characteristics the derivative solves
\begin{equation}
\label{eq:str5_a}
\frac{dv}{d\tb} = -2(1-u)v.
\end{equation}
We integrate (\ref{eq:str5_a}) using (\ref{eq:str4}) to find
\begin{equation}
\label{eq:str5}
v(\tb,s) =
\frac{v(\tb_0,s_0)e^{-2(\tb-\tb_0)}}{\left(1-u(\tb_0,s_0)(1-e^{-(\tb-\tb_0)})
  \right)^2}.  
\end{equation}
For $\tb \geq \tb_0$ we may take absolute values in (\ref{eq:str4})
and (\ref{eq:str5}) to obtain the decay estimates
\begin{equation}
\label{eq:str9}
|u(\tb,s)| \leq \frac{|u(\tau_0,s_0)| e^{-(\tb-\tb_0)}} {1- |u(\tb_0,s_0)|
 (1-e^{-(\tb-\tb_0)})}, 
\end{equation}
and 
\begin{equation}
\label{eq:str10}
|v(\tb,s)| \leq \frac{|v(\tb_0,s_0)| e^{-2(\tb-\tb_0)}}{\left(1 -
|u(\tb_0,s_0)|(1-e^{-(\tb-\tb_0)})\right)^2} \leq
\frac{|v(\tb_0,s_0)|e^{-2(\tb-\tb_0)}}{\left( 1 
-|u(\tb_0,s_0)|\right)^2} . 
\end{equation}

\subsection{Proof of Theorem~\ref{thm:strong_const}}
{\em 1.\/} We use the Fourier-Laplace inversion formula 
\begin{equation}
\label{eq:f_inv}
 \hx (\hn(\tb,\hx) - \nsg{0}(\hx)) = \frac{1}{2\pi}
  \int_{\mathbb{R}} e^{ik \hx} \left( v(\tb, ik) -
  v_{*,0}(ik)\right) \, dk.
\end{equation}
Thus, in order to prove (\ref{eq:strong_fourier}) it suffices to show 
\begin{equation}
\label{eq:strong_fourier2}
\lim_{\tb \to \infty} 
\int_{\mathbb R} |v(\tb,ik) - v_{*,0}(ik)| \,dk =0.
\end{equation}

{\em 2.\/} Let $\veps \in (0,\frac12)$ and put $R = \veps^{-1}$. 
We will prove (\ref{eq:strong_fourier2}) by estimating the integral
separately in three regions: $|k| \leq R$, $R \leq
|k| \leq Re^{\tb-T}$ and $Re^{\tb-T} \leq
|k|$, for $\tau\ge T$ where $T>0$ will be chosen sufficiently large,
depending on $\veps$ and the initial data $v_0$.
This is essentially the same decomposition used in the proof of
uniform convergence in 
the central limit theorem by Feller~\cite[XV.5.2]{Feller}. The main
new idea here is the use of the decay estimates (\ref{eq:str10}) and
the method of characteristics in the regions where $R \leq |k|$.

{\em 3.\/} $|k| \leq R$: Recall that the pointwise convergence of Laplace
transforms (\ref{eq:weak3}) is equivalent to 
$\hn(\tb,\hx) \,d\hx \to \hn_{*,0}(\hx)\,d\hx$ in the sense of weak
convergence of measures. Combined with (\ref{eq:mass_preserve}) this 
also implies that the mass measures  $\hx \hn(\tb,\hx) \,d\hx$
converge weakly to $\hx\hn_{*,0}(\hx)\,d\hx$ as $\tau \to \infty$. 
But this implies $v(\tb,ik)$ converges to $v_{*,0}(ik)$ uniformly  
for $|k| \leq R$~\cite[XV.3.2]{Feller}. 
Therefore,
\begin{equation}
\label{eq:new1}
\lim_{\tb \to \infty}  \int_{-R}^R |v(\tb,ik) - v_{*,0}(ik)| \,dk =0.
\end{equation}
{\em 4.} It remains to consider $|k| \geq R$.  It is 
sufficient to consider only $k \geq R$, since $|v(\tb,ik)|=|v(\tb,-ik)|$.
We will control $v(\tb,ik)$ and $v_{*,0}$ separately:
\[ \int_{R}^\infty |v(\tb,ik) - v_{*,0}(ik)| \,dk \leq \int_{R}^\infty
  |v(\tb,ik)| \,dk + \int_{R}^\infty |v_{*,0}(ik)| \,dk.\]
But $|v_{*,0}(ik)| = (1+|k|^2)^{-1}$ by (\ref{eq:weak2}), 
so that 
\[ \int_R^\infty |v_{*,0}(ik)| \,dk \leq R^{-1} = \veps.\]
In the rest of the proof we estimate $\int_R^\infty |v(\tb,ik)| dk$.

{\em 5.\/} Since 
$u(\tau,ik)\to u_{*,0}(ik)$ and $v(\tau,ik)\to v_{*,0}$ as
$\tau\to\infty$ for each real $k$, using (\ref{eq:weak3}) and (\ref{eq:weak2}) 
we may choose $T >0$ such that
\begin{equation}
\label{eq:str11}
 \sup_{\tb \geq T} |u(\tb,iR)| \leq R^{-1}= \veps, \quad
\sup_{\tb \geq T} |v(\tb,iR)| \leq R^{-2}. 
\end{equation}

{\em 6.\/} $R \leq k \leq Re^{\tb-T}$: 
The control obtained from (\ref{eq:str11}) propagates 
outwards along characteristics as $\tau$ increases. 
Precisely, whenever $\tb\ge T$, for any $k$ such that
$R \leq k \leq R e^{\tb-T} $ we have $ik=e^{\tb-\taus}iR$ where
$\taus\ge T$. By (\ref{eq:sol_s}) this means that
$ik = s(\tau;\taus,s_0)$ with $s_0=iR$. 
Then the decay estimate (\ref{eq:str10}) and the boundary control
(\ref{eq:str11}) imply
\begin{equation}
\label{eq:const_v_decay}
 |v(\tb,ik)| \leq \frac{|v(\taus,iR)| e^{-2(\tau- \taus)} }
{\left(1-|u(\taus, iR)|\right)^2}
 \leq \frac{1}{(1-\veps)^2} R^{-2} \left(\frac{R}{k}\right)^2 \leq 4 k^{-2}.
\end{equation}
Integrating this estimate we obtain
\[ 
\int_R^{{R}e^{\tb-T}} |v(\tb,ik)| \,dk \leq 
\int_R^\infty 4k^{-2} \,dk = 4 R^{-1} = 4 \veps.
\]

{\em 7.\/} ${R}e^{\tb-T} \leq k$: 
For brevity, let $\tilde{R}= Re^{-T}$. 
With $u_0(s):=u(0,s)$, $v_0(s):=v(0,s)$, 
we use (\ref{eq:str10}) and (\ref{eq:sol_s}) with $\taus=0$ to obtain
\begin{eqnarray}
\nonumber
\lefteqn{\int_{\tilde{R}e^\tb}^\infty |v(\tb,ik)| \,dk \leq e^{-2\tb}
  \int_{\tilde{R}e^\tb}^\infty
  \frac{|v_0(ike^{-\tb})|}{\left(1-|u_0(ike^{-\tb})|\right)^2} dk} \\ 
\nonumber
&& = e^{-\tb} \int_{\tilde{R}}^\infty \frac{
  |v_0(ik')|}{\left(1-|u_0(ik')|\right)^2} 
  dk' \leq \left( \sup_{|k'| \geq \tilde{R}}
    \frac{1}{\left(1-|u_0(ik')|\right)^2} \right) 
  e^{-\tb} \|v_0\|_{L^1}, 
\end{eqnarray}
where $k'= ke^{-\tau}$.
Since $|u_0(ik')| < 1$ for $k' \neq 0$ and $u_0(ik') \to 0$ as $k \to
\infty$ by the Riemann-Lebesgue lemma, we have  
$\sup_{|k'|\geq \tilde{R}} (1-|u_0(ik')|)^{-2} < \infty$.   

{\em 8.\/} Putting together the estimates we have obtained,
it follows that for $\tau$ sufficiently large, the integral in
\qref{eq:strong_fourier2} is less than $12\veps$.
This completes the proof.

\subsection{The discrete Smoluchowski equations}
We consider measure solutions of the form $\nu_t = \sum_{l=1}^\infty
n_l(t) \delta_{hl}(x)$, where $\delta_{hl}(x)$ denotes a Dirac mass at
$hl$. To avoid redundancy, we always assume that $h$ is the {\em
  span\/} of the lattice, that is, the maximal $h>0$ so that all
initial clusters, and thus clusters at any time $t>0$, are
concentrated on $h\mathbb{N}$. We will call $\nu_t$ a lattice measure
with span $h$.  
Notice that if the initial number of clusters and the mass are
finite, by rescaling $n_l$ and $h$ we may
assume that $\int_0^\infty \nu_1(dx) =
\int_0^\infty x \nu_1(dx) =1$. Under these conditions, the weak
convergence theorem of~\cite{MP1} asserts that 
$\lim_{\ta \rightarrow \infty} \ta \etalap(t,s/\ta)
=u_{*,0}(s)$. We show that this theorem 
may be strengthened by use of Fourier series.
The Fourier transform of $\nu_t$ is the Fourier series
\[
\etalap(\ta,ik) = \sum_{l\in\N} n_l(\ta)e^{-ilhk}, \ k\in\R,
\]
which has minimal period $2\pi/h$. Thus 
$n_l(\ta) = ({h}/{2\pi}) \int_{-\pi/h}^{\pi/h} e^{ilhk}
\etalap(\ta,ik)\,dk$, or
\begin{equation}
\ta^2 n_l(\ta) = \frac{h}{2\pi} \int_{-\pi e^\tb/h}^{\pi e^\tb/h} 
\exp(ilhke^{-\tb}) u(\tb,ik)\,dk,
\end{equation}
in similarity  variables from~(\ref{eq:str1}). We integrate by parts
and let 
\begin{equation}\label{eq:disc-scale}
\hat{l}= lhe^{-\tb} = lh\ta^{-1},\quad \hn_l(\ta) =
h^{-1}\ta^2n_l(\ta)
\end{equation}
to obtain
\begin{equation}
\label{eq:discrete_inv3}
\hat{l} \hn_{l}(\ta) =\ta l n_l(\ta) 
= \frac{1}{2\pi} \int_{-\pi
e^\tb/h}^{\pi e^\tb/h}  e^{i  \hat{l} k} v(\tb,ik)\,dk. 
\end{equation}
As in Theorem~\ref{thm:strong_const} we expect the right hand side to
converge to $\hat{l}\nsg{0}(\hat{l})$ as $\tau \to\infty$, indeed
uniformly for $\hat{l}\in ht^{-1}\N$.  
\begin{thm}
\label{thm:const_discrete}
Let $\nu_1 \geq 0$ be a lattice measure with span $h$ such that
$\int_0^\infty \nu_1(dx) = \int_0^\infty x \nu_1(dx)=1$. 
 Then with the scaling (\ref{eq:disc-scale}) we have
\begin{equation}
\label{eq:const_lim}
\lim_{\ta \rightarrow \infty}\, \sup_{l\in\N}\,
\hat{l} \left| \hn_{l}(\ta)
- \nsg{0}(\hat{l}) \right|=0.
\end{equation}
\end{thm}
\begin{proof}
By (\ref{eq:discrete_inv3}) and the  continuous Fourier inversion
formulas it suffices to show that 
\[ \lim_{\tb \rightarrow \infty}\sup_{\hat{l}\ge0}
\left| \int_{-\pi e^\tb/h}^{\pi e^\tb/h}
  e^{i  \hat{l} k} v(\tb,ik) \,dk 
  - \int_{\mathbb{R}} e^{i\hat{l} k} v_{*,0}(ik) \,dk \right| =0.\]
As earlier it suffices to consider $k >0$. The integrals
\[ \int_{-R}^R | v(\tb,ik) - v_{*,0}(ik)| \,dk, \quad
  \int_R^{\tilde{R}e^\tb} |v(\tb,ik)|\,dk,  \quad 
  \int_R^\infty |v_{*,0}(ik)| \,dk,\]
with $\tilde{R}=Re^{-T}$, are controlled exactly as 
in the proof of Theorem~\ref{thm:strong_const}.
It only remains to estimate the integral of $|v(\tau,ik)|$ over the region
  $\tilde{R}e^\tb < k < \pi e^\tb/h$. We assume that $\pi/h > \tilde{R}$,
for  otherwise there is nothing to prove. But then by 
the formula (\ref{eq:sol_s}), the uniform decay estimate
  (\ref{eq:str10}), and the change of variables $k' = k e^{-\tau}$,  we have
\[ 
\int_{\tilde{R}e^\tb}^{\pi e^\tb/h} |v(\tb,ik)| \,dk \leq e^{-\tb}
  \int_{\tilde{R}}^{\pi/h} \frac{|v_0(ik')|}{\left| 1-u_0(ik')(1-e^{-\tau}) 
  \right|^2} dk' .
  \]
Since the domain of integration is finite, it suffices to show that
the integrand is uniformly bounded in time. Since $|v_0(ik)| \leq 1$, 
it is only necessary to control the denominator. 
But $u_0(ik) =  \sum_{l\in\N} n_l(0)e^{-ilkh}$ with $n_l(0) \geq 0$.
Therefore, $|u_0(ik)| \leq 1$, and~\cite[XV.1.4]{Feller} yields that
\[u_0(ik)=1 \quad\mbox{if and only if}\quad k= \frac{2\pi m}{h}, \ m \in
\Z.\] In particular, we have the strict inequality \[ \min_{k \in
[\tilde{R},\frac{\pi}{h}]} \left|1-u_0(ik) \right| \geq \delta >0.\]
Therefore, \[ \left|1 -u_0(ik)(1-e^{-\tau}) \right|   \geq |1-u_0(ik)|
- |u_0(ik)| e^{-\tau} \geq \delta - e^{-\tau} \geq \frac{\delta}{2} \]
for sufficiently large $\tau$. Thus,
\[
\int_{\tilde{R}e^\tb}^{\pi e^\tb/h} |v(\tb,ik)| \,dk \leq 
\frac{2\pi}{\delta h}e^{-\tb}.
\]
\end{proof}

\section{Uniform convergence of densities for the additive kernel}
\subsection{Rescaling and approach to self-similarity}
In this section we prove the analogs of
Theorems~\ref{thm:strong_const} and ~\ref{thm:const_discrete}
for the additive kernel. The essential geometric ideas of the proof are 
similar to the previous section. However, the trajectories
of the characteristic curves $s(t;t_0,s_0)$ in the complex plane 
are no longer rays, 
and the proofs require more careful analysis. 
As earlier, we will work with the explicit solution
formula for an appropriate Laplace transform. For $z \in \Cclose$ we define 
\begin{equation}
\label{eq:phi_def}
\etaadd(t,z) = \int_0^\infty \left(1 - e^{-zx} \right) n(t,x) dx.
\end{equation}
We observe that $1-e^{-zx} = zx + O(z^2x^2)$ as $x \to 0$. 
We use $\etaadd$ instead of the standard Laplace transform of $n$
because the latter may not be well-defined:
E.g., the similarity profile $\nsg{1}$ 
in (\ref{eq:ndef}) satisfies $\nsg{1}(x)\sim Cx^{-3/2}$ as $x \to 0$. 
More generally, one needs the initial data to have only a finite 
first moment for existence and
uniqueness of a solution to (\ref{eq:smol1}) in the case of the
additive kernel~\cite{MP1}.
A deeper reason for this choice of variables (and notation) is probabilistic:
(\ref{eq:phi_def}) is the \LK\/ formula for the Laplace exponent of a
subordinator with no drift~\cite{Bertoin}. We will always assume that
the initial data $n_0$ satisfies the moment conditions
\begin{equation}
\la{eq:moment_add}
\int_0^\infty x n_0(x) dx = 1, \quad \int_0^\infty x^2 n_0(x) dx =1.  
\end{equation}
We substitute (\ref{eq:phi_def}) in (\ref{eq:smol1}) and use
(\ref{eq:moment_add}) to see that $\Phi(t,z)$ solves the equation
\begin{equation}
\label{eq:phi_burgers}
\partial_t \etaadd -\etaadd \partial_{z} \etaadd = -\etaadd, \quad \etaadd(0,z)=
\int_0^\infty(1-e^{-zx}) n_0(x) dx.
\end{equation}

As shown in \cite{MP1} by the method of characteristics,
\qref{eq:phi_burgers} has a unique solution for $z>0$, $t>0$ which is
analytic with derivative $\D_z\etaadd$ completely monotone in $z$ and
satisfying $\D_z\etaadd(t,0)=1$ for all $t$.  For each $t>0$ then,
$\D_z\etaadd(t,\cdot)$ is the Laplace transform of a probability
measure, so its domain contains $\Cclose$ and \qref{eq:phi_burgers} holds
by analytic continuation for $z\in \Cint$, $t>0$.

In  contrast with (\ref{eq:const_sol}) it is
not obvious that a suitable rescaling will lead to convergence to
self-similar form. This point is discussed in~\cite[Sec.7]{MP1}, and
we refer the reader to that article for motivation for the following
change of variables. We define the similarity variables
\begin{equation}
\la{eq:add_rsc1}
 \hx = x e^{-2t}, \quad s = z e^{2t},
\end{equation}
and the rescaled number density
\begin{equation}
\la{eq:add_rsc2}
\quad \hn(t, \hx) = e^{4t} n(t,
\hx e^{2t}) = e^{4t} n(t, x).
\end{equation}
We also define the rescaled Laplace transforms
\begin{equation}
\la{eq:add_rescaling1}
\varphi(t,s) = e^{2t} \etaadd(t,e^{-2t}s) = \int_0^\infty (1-e^{-s\hx}) \hn(t,
\hx) d\hx. 
\end{equation}
Part of the motivation for the  rescaling (\ref{eq:add_rsc1}) and
(\ref{eq:add_rsc2}) is that this choice preserves
{\em both} moment conditions in (\ref{eq:moment_add}). That is, we have 
\begin{equation}
\la{eq:moment_add2}
\int_0^\infty \hx \hn(t,\hx) d\hx = \int_0^\infty
\hx^2 \hn(t, \hx) d\hx =1, \quad t \geq 0.
\end{equation}
This should be compared with (\ref{eq:mass_preserve}) for the constant
kernel. The mass measure  plays the same role here as the number 
measure did for  $K=2$. Thus, we denote its Laplace transform by
the same letter, and let
\begin{equation}
\la{eq:add_rescaling_u}
u(t,s) = \partial_s \vp(t,s) = 
\int_0^\infty e^{-s\hx} \hx \hn(t,\hx) d \hx.
\end{equation}
By Theorem 7.1 in~\cite{MP1} (also see \cite[Appendix G]{Leyvraz}), the 
assumptions in~(\ref{eq:moment_add}) imply that the rescaled mass measures
converge to the similarity profile, with
\begin{equation}
\la{eq:add_weak}
\hx \hn(t,\hx) d \hx \to \hx \nsg{1}(\hx) d\hx = \frac{1}{\sqrt{2\pi}}
  \hx^{-1/2} e^{-\hx/2} d\hx, \quad t \to \infty
\end{equation}
in the sense of weak convergence of measures. 
It then follows from~\cite[XIII.1.2]{Feller} that 
(\ref{eq:add_weak}) is equivalent to
\begin{equation}
\la{eq:add_weak1}
\lim_{t \to \infty} u(t,s) = \frac{1}{\sqrt{1+2s}} =: u_{*,1}(s). 
\quad s \in \Cclose,
\end{equation}

Our goal is to strengthen (\ref{eq:add_weak}) to uniform convergence of 
densities for (\ref{eq:smol1}) and uniform convergence of coefficients
for (\ref{eq:disc-smol}). For the continuous Smoluchowski equations we
prove 
\begin{thm}
\label{thm:strong_add}
Suppose $n_0(x) \geq 0$,
$\int_0^\infty xn_0(x) dx = \int_0^\infty x^2 n_0(x) dx =
1$. Suppose also that the Fourier transform of $x^2n_0$ 
is integrable. Then in terms of the rescaling
(\ref{eq:add_rsc1})--(\ref{eq:add_rsc2}) we have 
\begin{equation}
\label{eq:strong_fourier_add}
\lim_{t \to \infty}\, \sup_{\hx>0}
\,\hx^2| \hn(t,\hx) - \nsg1(\hx) |  =0,
\end{equation}
where $\nsg1(\hx)$ is the similarity profile defined in (\ref{eq:ndef}).
\end{thm}
Once Theorem~\ref{thm:strong_add} is established, it is relatively
straightforward to obtain the analogous result for the discrete
Smoluchowski equations; see Theorem~\ref{thm:add_discrete} below. Thus, most of
our effort is devoted to Theorem~\ref{thm:strong_add}.

Observe that we prove uniform convergence of
the weighted density $\hx^2 \hn(t,\hx)$. 
As in the previous section, this is because
Theorem~\ref{thm:strong_add} is proved 
using the Fourier-Laplace inversion formula. 
Since $|u_{*,1}(ik)| \sim |k|^{-1/2}$ as 
$|k| \to \infty$,  $u_{*,1}$ is not integrable on the imaginary
axis. This divergence is 
due to the fact that $\nsg{1}(\hx) =0$ for $\hx <0$ and
$\hx \nsg{1}(\hx) \sim C\hx^{-1/2}$ as $\hx\to0^+$. 
As earlier, we resolve the situation by considering 
the transform of the next moment. Let
\begin{equation}
\la{eq:def_v}
v (t,s) = -\partial_s u(t,s) = \int_0^\infty e^{-s\hx}
\hx^2 \hn(t,\hx) d\hx, \quad s \in \Cclose .
\end{equation}
We integrate and differentiate (\ref{eq:add_weak1}) to obtain 
\begin{equation}
\la{eq:def_v*}
\varphi_{*,1}(s) = \sqrt{1+2s}-1, \quad v_{*,1}(s) = (1+2s)^{-3/2},
\quad s\in \Cclose.
\end{equation}

\subsection{Characteristics and estimates}
The equations of evolution for $\varphi$ and $u$ are
\begin{eqnarray}
\label{eq:evol_phi}
\partial_t \varphi + (2s- \varphi) \partial_s \varphi &=& \varphi,\\
\label{eq:add_str2}
\partial_t u + (2s- \varphi) \partial_s u & =& -u(1-u).
\end{eqnarray}

In what follows, we first derive solution formulas to
(\ref{eq:evol_phi}) by the method of characteristics. 
We then show that the solution map for the characteristic equation
is never degenerate and
that characteristics flow out of the right 
half into the left half of the complex plane as $t$ increases.
For most parts of our analysis, it will suffice to study characteristics 
in the right half plane only. But for one part, 
we need to study characteristics that start in the right half plane
but move into the left half plane. 

We use the notation $s(t;t_0,s_0)$ to denote the solution to 
\begin{equation}
\label{eq:add_str3}
\frac{ds}{dt} = 2s -\varphi, \quad s(t_0;t_0,s_0) =s_0.
\end{equation}
Along the characteristic curve $s(t;t_0,s_0)$ we have
\begin{equation}
\label{eq:add_str4}
\frac{d\varphi}{dt} = \varphi, \quad \mathrm{and} \quad \frac{du}{dt}
= -u(1-u). 
\end{equation}
We integrate (\ref{eq:add_str4}) to obtain 
\begin{equation}
\label{eq:u_soln}
\varphi(t,s) = e^{t-t_0} \varphi (t_0,s_0), \quad
u(t,s) = \frac{u(t_0,s_0)e^{-(t-t_0)}}{1- u(t_0,s_0)(1-e^{-(t-t_0)})}.
\end{equation} 
We now substitute for $\varphi(t,s)$ from (\ref{eq:u_soln}) in
(\ref{eq:add_str3}) and integrate to obtain the explicit solution 
\begin{equation}
\label{eq:char_soln}
e^{-2(t-t_0)} s(t;t_0,s_0) = s_0 - \varphi(t_0,s_0)(1-e^{-(t-t_0)}).
\end{equation}
This equation can also be rewritten in two other useful forms, namely
\begin{equation}
\label{eq:phi_time2}
e^{-2(t-t_0)} \left( s - \vp(t,s) \right) = \left(s_0 - \vp(t_0,s_0)
\right),
\end{equation}
and
\begin{equation}\label{phis}
\frac{\varphi(t,s)}{s} = \frac
{(\vp(t_0,s_0)/s_0) e^{-(t-t_0)} }
{1- (\vp(t_0,s_0)/s_0)(1-e^{-(t-t_0)}) } .
\end{equation}
The method of characteristics also yields an explicit solution for 
$v(t,s)$. We differentiate equation (\ref{eq:add_str2}) to  obtain 
\begin{equation}
\label{eq:add_str5}
\frac{dv}{dt} = -3(1-u) v.
\end{equation}
We substitute for $u$ from (\ref{eq:u_soln}) and integrate
(\ref{eq:add_str5}) to obtain
\begin{equation}
\label{eq:v_soln}
v(t,s) = \frac{v(t_0,s_0)e^{-3(t-t_0)}}
{\left(1- u(t_0,s_0)(1-e^{-(t-t_0)})\right)^3}. 
\end{equation}

Let $\vp_0(s):=\vp(0,s)$ and similarly $u_0(s):=u(0,s)$, 
$v_0(s):=v(0,s)$.
Since $u = \partial_s \vp$ and $\varphi(t,0)=0$, 
the  moment conditions (\ref{eq:moment_add}) 
and the identity $\vp_0(s)/s=\int_0^1 u_0(\tau s)\,d\tau$ imply 
\begin{equation}\label{u0est}
|u_0(s)|\le1, \quad |v_0(s)|\le 1, \quad |\vp_0(s)|\le |s|, \
s\in\Cclose.
\end{equation}
These inequalities are strict for $s\ne0$ because $xn_0(x)\,dx$ is not
a lattice measure \cite[XV.1.4]{Feller}.
Taking $t_0=0$ at first, 
for $t \geq t_0$ we take absolute values in  
(\ref{eq:u_soln}) and (\ref{eq:v_soln})
to see that $|u|$ and $|v|$ decay along characteristics according to
\begin{equation}
\label{eq:add_u_decay}
|u(t,s)| \leq \frac{|u(t_0,s_0)|e^{-(t-t_0)}}{ 1
 -|u(t_0,s_0)|(1-e^{-(t-t_0)})},  
\end{equation}
\begin{equation}
\label{eq:add_str10}
|v(t,s)| 
\leq  \frac{|v(t_0,s_0)|e^{-3(t-t_0)}}{\left( 1  -|u(t_0,s_0)|\right)^3} . 
\end{equation}

From \qref{eq:add_u_decay} and the fact that $|u_0(s_0)|<1$ for
$s_0\ne0$, and a similar estimate using \qref{phis} and
$|\vp_0(s_0)/s_0|<1$, it follows 
\begin{equation}
\la{eq:uniform_u}
|u(t,s)| < 1, 
\quad \left|{\varphi(t,s)}/{s}\right| <1, 
\quad t \geq 0, \ s\ne0 .
\end{equation}
Then \qref{eq:add_u_decay} and \qref{eq:add_str10} hold also 
for any $t_0\ge0$, if $t\ge t_0$.

Let us also note the uniform outward growth of characteristics implied
by (\ref{eq:uniform_u}). Using (\ref{eq:uniform_u}) together with
(\ref{eq:add_str3}) we obtain
\begin{equation}
\la{eq:uniform_s}
|s| \leq \frac{d|s|}{dt} \leq 3|s|.
\end{equation}
Thus, $|s_0|e^{(t-t_0)} \leq |s| \leq e^{3(t-t_0)}|s_0|$. We will refine
this crude estimate in the proof of Theorem~\ref{thm:strong_add}, but
we note here that $|s(t;t_0,s_0)|$ is a strictly increasing function of
$t$. 

In addition to the decay along characteristics, 
we will need the following uniform
Riemann-Lebesgue lemma. Let 
$C_R =\{s \in \Cclose \mid |s| =R\}$
denote the semicircle of radius  $R$ in the right half plane.
\begin{lemma}
\label{le:riemann-lebesgue}
Let $g(x) \in L^1(0,\infty)$  and $G(s) = \int_0^\infty
  e^{-sx} g(x) dx$. Then 
\begin{equation}
\label{eq:riemann-lebesgue}
\lim_{R \rightarrow   \infty} \sup_{s\in C_R}|G(s)| =0.
\end{equation}
\end{lemma}
\begin{proof}
Let $\varepsilon >0$. 
We choose a step function $g_\varepsilon =
\sum_{k=1}^{K} c_{k} \mathsf{1}_{[a_k,b_k]}$ so that
$\|g -g_\varepsilon\|_{L^1} < \varepsilon$. But then,
$\|e^{-sx}(g -g_\varepsilon)\|_{L^1} < \varepsilon$. 
Therefore, for $s \in \Cclose$, 
\[|G(s)| \leq \varepsilon + \left| \int_0^\infty e^{-sx}
    g_\varepsilon(x) 
    dx \right| 
= \varepsilon + 
\left| \sum_{k=1}^{K} c_{k} \int_{a_k}^{b_k}
  e^{-sx}dx 
\right| \leq \varepsilon + \frac{C_\varepsilon}{|s|}.\]
\end{proof}
We apply this lemma and (\ref{eq:moment_add2}) 
to $g(\hx)=\hx^j \hn(t,\hx)$ for $j=1, 2$ to infer that
for every $t \geq 0$, as $|s|\to\infty$ with $\Re s\ge0$ we have 
\begin{equation}\label{udecay}
|u(t,s)|\to 0, \quad
|v(t,s)|\to 0, \quad
\left|\frac{\vp(t,s)}s\right|\to0.
\end{equation}

\subsection{Geometry of the characteristic map in the complex plane}
\nwc{\Dbar}{\bar{\Omega}}
\nwc{\dom}{\Omega}
In this subsection, we  study the solution formula (\ref{eq:char_soln}).
Our goal is to delineate some key properties of the map 
$s_0\mapsto s(t;t_0,s_0)$ for $t, t_0\ge0$.

Let $\Cint$ denote the open right half plane. We let $\dom_t$
denote the image of $\Cint$ under the map $s_0\mapsto s(t;0,s_0)$,
and let $\Gamma_t$ denote the image of the imaginary axis under the
same map. We aim to prove the following.

\begin{lemma}\label{le:char_map}
\begin{itemize}
\item[(i)] For any $t>0$, $\Gamma_t$ is a $C^2$ curve that 
passes through the origin but otherwise 
lies in the open left half plane. On $\Gamma_t$, 
$\Re s$ is a $C^2$ function of $\Im s$.
\item[(ii)] $\dom_t$ is the component of the complex plane to the right of
$\Gamma_t$. 
Consequently $\Gamma_t=\D\dom_t$ and $\dom_t\supset \Cclose\setminus\{0\}$.
\item[(iii)] Whenever $t_1\ge t_0\ge0$, the map 
$s_0\mapsto s_1=s(t_1;t_0,s_0)$ is
one to one from $\Dbar_{t_0}$ onto $\Dbar_{t_1}$.
It is $C^2$ on $\Dbar_{t_0}$ and analytic in $\dom_{t_0}$.
The inverse map
is given by $s_1\mapsto s_0=s(t_0;t_1,s_1)$, and  
is $C^2$ on $\Dbar_{t_1}$ and analytic in $\dom_{t_1}$.
\item[(iv)] Whenever $t_1\ge0$ and $s_1\in\Cclose$, the backward
characteristic curve $s(t_0;t_1,s_1)$, $t_0\in[0,t_1]$, lies in $\Cclose$.
\end{itemize}
\end{lemma}

\begin{proof}
We first establish part (iii), taking $t_0=0$ at first. 
Since $x^2n_0$ is integrable, $v_0(s)$ is continuous in $\Cclose$ and
analytic for $\Re s>0$. 
It follows by a standard dominated convergence argument that
$u_0$ is $C^1$ and $\vp_0$ is $C^2$ in $\Cclose$, and these functions are
analytic in $\Cint$. From \qref{eq:char_soln} we see that the map
$s_0\mapsto s(t;0,s_0)$ is analytic in $\Cint$ and $C^2$ on $\Cclose$ 
(meaning derivatives up to second order extend continuously to $\Cclose$).

We next claim that this map is one to one. The proof relies on 
the fact that $\vp_0$ is contractive, with
\begin{equation}
\label{eq:Lip_vp}
|\vp_0(\tilde{s}_0) - \vp_0(s_0) | \leq |\tilde{s}_0 - s_0|, \quad
\tilde{s}_0,s_0, \in \Cclose.
\end{equation}
This holds because $| \partial_s \vp_0(s)| \leq
1$ for $s\in\Cclose$ as an immediate consequence of
(\ref{eq:moment_add2}) and (\ref{eq:add_rescaling_u}). 
Now suppose $s(t;0,\tilde{s}_0)=s(t;0,s_0)$ where $\tilde{s}_0, s_0
\in \Cclose$. 
Then (\ref{eq:char_soln}) implies 
\[ \tilde{s}_0 -s_0 = \left( 1-e^{-t} \right) \left( \vp_0(\tilde{s}_0)-
  \vp_0(s_0) \right). \]
From this and \qref{eq:Lip_vp} we infer

We observe that the derivative of this map is uniformly bounded away
from zero. Indeed, \qref{eq:char_soln} and \qref{u0est} yield
\[
\left| \frac{d s}{d s_0} \right| 
\ge e^{2t}\left( 1- |u_0(s_0)|(1-e^{-t})\right)
\ge e^t.
\]
It follows by the inverse function theorem that $\dom_t$ is an open set,
and by continuity the image of $\Cclose$ is $\Dbar_t$.  The inverse map from
$\Dbar_t$ to $\Cclose$ is analytic in $\dom_t$, and $C^2$ on
$\Dbar_t$.

For $t_1>0$, the inverse of the map $s_0\mapsto s_1=s(t_1;0,s_0)$ may
be obtained by solving the characteristic equation in
\qref{eq:add_str3} backwards from time $t_1$ to $t_0=0$, so we have
$s_0=s(0;t_1,s_1)$. Now whenever $t_1\ge t_0\ge0$ in general,
we may follow any characteristic curve back from a point in
$\Dbar_{t_1}$ at time $t_1$ to a point in $\Cclose$ at time $0$ and then
forward to a point in $\Dbar_{t_0}$ at time $t_0$. This means that
$s(t_1;t_0,s_0)=s(t_1;0,s(0;t_0,s_0))$.
Part (iii) of the lemma now follows from the properties established
in the case $t_0=0$.

Next we prove part (i). For $t>0$, $\Gamma_t$ is the image of the map
$k\mapsto s(t;0,ik)= e^{2t}(ik-\vp_0(ik)(1-e^{-t}))$, $k\in\R$, 
and this is a $C^2$ function of $k$. 
We have $s(t;0,0)=0$, but $\Re s<0$ for $k\ne0$.
This is so because $\Re s$ and $\Re\vp_0(ik)$ have opposite signs,
and
\[ \Re \vp_0(ik) = \int_0^\infty (1-\cos kx) n_0(x) dx > 0,
\quad k\neq 0, \] 
since $n_0$ is continuous. Finally, we find that 
\[
\Im \frac{d}{dk} s(t;0,ik) \ge e^{2t}(1- |u_0(ik)|(1-e^{-t})) >0
\]
using \qref{u0est}. Hence $\Re s$ is a function of $\Im s$ on
$\Gamma_t$.

Now we establish part (ii). 
By \qref{udecay} we have that as $|s_0|\to\infty$ with
$s_0\in\Cclose$, $|\vp_0(s_0)/s_0|\to 0$, so $s=s_0 e^{2t}(1+o(1))$ by
\qref{eq:char_soln}. Let $s_1\in\C$ lie to the right of $\Gamma_t$,
and put $f(s_0)= s(t;0,s_0)-s_1$. 
It follows by applying the argument principle to large semicircles
that the analytic function $f$ has a single zero at some point $s_0\in\Cint$. 
Indeed, $\arg f(re^{i\theta})\to\theta$ as $r\to\infty$ for 
$-\frac\pi2\le \theta\le \frac\pi2$, and as $k$ goes from $\infty$ to
$-\infty$, $f(ik)$ does not cross the positive real axis, 
so $\arg f(ik)$ changes from $\frac\pi2$ to $\frac{3\pi}2$.
Thus, $f$ maps a large semicircle to a curve that winds exactly once
about 0.  Hence $s_1\in\dom_t$.

Finally, part (iv) follows by a change of variables, replacing $t-t_0$
by $t$, and applying parts (i)--(iii).  
\end{proof}

\subsection{Proof of Theorem~\ref{thm:strong_add}}
{\em 1.\/} By the Fourier-Laplace inversion formula, 
it suffices to prove
\begin{equation}\label{eq:add-lim}
\lim_{t \to \infty} \sup_{x>0} 
\left| \int_{\mathbb{R}} 
e^{ikx} \left[ v(t,ik) - v_{*,1}(ik)\right] \,dk \;\right| =0.
\end{equation}

{\em 2.\/} Let $\veps \in (0,\frac18)$, and put $R = \frac12 \veps^{-2}$. 
We will prove (\ref{eq:add-lim}) by estimating the integral
for $t\ge T$ separately in three regions: $|k| \leq R$, 
$R \leq |k| \leq \tilde{R}e^{2t}$ and $\tilde{R}e^{2t} \leq |k|$,
where $\tilde{R}=Re^{-2T}$ and $T$ depends only on $\veps$ and 
the initial data $v_0$.
This is the same decomposition used in the proof of
Theorem~\ref{thm:strong_const}, and convergence in the region $|k|
\leq R$ will follow as before. However, estimates for $|k|
\geq R$ are more subtle, and use the analyticity and geometry of the
characteristic map. 

{\em 3.\/} $|k| \leq R$: Theorem 7.1 in~\cite{MP1} implies that 
$\hx \hn(\tb,\hx) \,d\hx \to \hx \hn_{*,0}(\hx)\,d\hx$ in the sense of weak
convergence of measures. Combined with (\ref{eq:moment_add2}) this 
also implies that the measures  $\hx^2 \hn(\tb,\hx) \,d\hx$
converge weakly to $\hx^2\hn_{*,1}(\hx)\,d\hx$ as $t \to \infty$. 
But this implies $v(t,ik)$ converges to $v_{*,1}(ik)$ uniformly  
on compact subsets of $\Cclose$, and in particular on compact subsets of
the imaginary axis~\cite[XV.3.2]{Feller}. Thus,
\begin{equation}
\label{eq:add_new1}
\lim_{t \to \infty}  \int_{-R}^R |v(t,ik) - v_{*,1}(ik)| \,dk =0.
\end{equation}

{\em 4.} $|k| \geq R$:  It is 
sufficient to consider only $k \geq R$, since $|v(t,ik)|=|v(t,-ik)|$.
We will control $v(t,ik)$ and $v_{*,1}$ separately:
\[ \int_{R}^\infty |v(t,ik) - v_{*,1}(ik)| \,dk \leq \int_{R}^\infty
  |v(t,ik)| \,dk + \int_{R}^\infty |v_{*,1}(ik)| \,dk.\]
But $|v_{*,1}(ik)| \leq (2k)^{-3/2}$ by
(\ref{eq:def_v*}). Thus, 
\[ \int_R^\infty |v_{*,1}(ik)| \,dk \leq \int_R^\infty (2k)^{-3/2}
\,dk  = (2R)^{-1/2} = \veps.\]

{\em 5.\/} In the rest of the proof we estimate $\int_R^\infty |v(t,ik)|
dk$. In order to aid the reader, we state the main estimates as two
distinct lemmas.
\begin{lemma}
\la{le:transition}
Let $\veps \in (0,\frac18)$. There exists $T >0$ 
depending on $\veps$ and the 
initial data, and a universal constant $C$, such that if $t \geq T$ then
\begin{equation}
\la{eq:addnew2}
\int_R^{Re^{2(t-T)}} |v(t,ik)| \, dk \leq C\veps.
\end{equation}
\end{lemma}
\begin{lemma}
\la{le:contour_decay}
Let $\tilde{R} >0$. There exists $\tilde{C}$ depending on
$\tilde{R}$  
and the initial data such that for all $t \geq 0$ we have
\begin{equation}
\la{eq:addnew3}
\int_{\tilde{R}e^{2t}}^\infty |v(t,ik)| \, dk \leq \tilde{C} e^{-t}.
\end{equation}
\end{lemma}
{\em 6.\/} We now prove (\ref{eq:add-lim}). We choose $T$ as in
Lemma~\ref{le:transition}, and then $\tilde{R}= Re^{-2T}$ in
Lemma~\ref{le:contour_decay}.  
Choose $T_* \geq T$ such that for $t \geq T_*$ 
\[ \int_{-R}^R |v(t,ik)-v_{*,1}(ik)| \, dk < \veps,\qquad \tilde{C}
e^{-t} \leq \tilde{C} e^{-T_*} < \veps. \]
Thus, for $t \geq T_*$ we have 
\ba
\nn
\lefteqn{\int_{\mathbb{R}} |v(t,ik)-v_{*,1}(ik)| \, dk \leq
  \int_{-R}^R |v(t,ik)-v_{*,1}(ik)| \, dk}\\ 
\nn
&& + 2 \left( \int_R^\infty |v_{*,1}(ik)|\,dk +  \int_R^{\tilde{R}e^{2t}}
  |v(t,ik)|\, dk 
+  \int_{\tilde{R}e^{2t}}^\infty |v(t,ik)|\, dk \right)\\
\nn 
&& \leq \veps + 2 \left( \veps + C\veps + \veps\right).
\ea 
Since $\veps\in(0,\frac18)$ may be chosen arbitrarily small, this
completes the proof.
   
\subsection{Proof of Lemma~\ref{le:transition}}
In this subsection we will always suppose $s\in \Cclose$. 
In a manner similar to step 6 of the proof of Theorem~\ref{thm:strong_const},
the idea is to get estimates on the semicircle
$C_R :=\{s \in \Cclose \mid |s|=R\}$ valid for large time, and propagate
these estimates outwards along characteristics.  We first use
(\ref{eq:add_weak1}) and (\ref{eq:def_v*}) to obtain 
the following estimates for $s \in \Cclose$:
\begin{equation}
\la{eq:def_v**}
|\vp_{*,1}(s)| < |2s|^{1/2}, \quad |u_{*,1}(s)| < |2s|^{-1/2}, \quad
|v_{*,1}(s)| < |2s|^{-3/2}. 
\end{equation}
Next, we use the uniform convergence
on compact sets and (\ref{eq:def_v**}) to see that there exists $T_0$ 
(depending on $\veps$ and the initial data) such
that for all $s_0 \in C_R$ and $t_0 \geq T_0$ we have
\begin{eqnarray}
\label{eq:decay_u_add1}
|\vp(t_0,s_0)/s_0| & \leq & 2 (2R)^{-1/2} = 2\veps \leq 1/4,  \\
\label{eq:decay_u_add2}
|u(t_0,s_0)| & \leq & (2R)^{-1/2} = \veps, \\
\label{eq:decay_u_add3}
|v(t_0,s_0)| & \leq & (2R)^{-3/2} = \veps^3.
\end{eqnarray}
We first extend (\ref{eq:decay_u_add1}) to a larger domain in $s$. 

\noindent
{\em Claim 1:\/} There exists $T_1 \geq T_0$ such that 
\begin{equation}
\la{eq:decay_u_add4}
\left|\frac{\vp(t,s)}s\right| \leq  1/3, 
\quad t \geq T_1,\ s\in\Cclose,\ |s| \geq R. 
\end{equation}
{\em Proof of claim 1:}
Observe that by using (\ref{eq:uniform_u}) and (\ref{udecay}) in
\qref{phis}, we have 
\[
\att:= \sup\{|\vp(T_0,s)/s| \mid 
s\in\Cclose,\ |s|\ge R \} < 1. 
\]
Fix $t_1\ge T_0$, $s_1\in\Cclose$ with $|s_1|\ge R$. 
Either the characteristic curve $s(t;t_1,s_1)$ that passes through
$s_1$ at time $t_1$ intersects $C_R$ at some time $t_0\in [T_0,t_1]$, 
or not.
If so, then  
$s_1=s(t_1;t_0,s_0)$ for some $s_0\in C_R$, and 
(\ref{phis}) and \qref{eq:decay_u_add1} directly yield
\[
\left|\frac{\vp(t_1,s_1)}{s_1}\right|
\le \frac{1/4}{1-1/4} = \frac{1}{3}.
\]
If not, then $|s(t;t_1,s_1)|>R$ for all $t\in[T_0,t_1]$, by continuity
and the fact that $s(t;t_1,s_1)\in\Cclose$ for all $t\in[0,t_1]$ by
part (iv) of Lemma~\ref{le:char_map}.
Then taking $t_0=T_0$, $s_0=s(T_0;t_1,s_1)$ in \qref{phis} yields 
\[
\left|\frac{\vp(t_1,s_1)}{s_1}\right|
\le \frac{\att e^{-(t_1-T_0)}}{1-\att}\le \frac{1}{3},
\]
provided $t_1 \geq T_1$ with $T_1$ sufficiently large. This proves the claim.

\noindent
{\em Claim 2:\/} Let $T=T_1+\frac12\ln 2$.
Suppose $t_1 \geq T$ and $ R \leq |s_1| \leq Re^{2(t_1-T)}$.
Then  the characteristic curve $s(t;t_1,s_1)$ that passes through
$s_1$ at time $t_1$ intersects $C_R$ at some time $t_0\in [T_1,t_1]$. 

{\em Proof of claim 2:}. Suppose the claim were false. Then the
continuity of $|s(t;t_1,s_1)|$ and part (iv) of
Lemma~\ref{le:char_map} imply $R <|s(t_0;t_1,s_1)| $ for 
all $t_0 \in [T_1,t_1]$. But now, by \qref{eq:phi_time2} with
$s_0=s(t_0;t_1,s_1)$ we have
\begin{equation}\label{s0est}
s_0\left(1-\frac{\vp(t_0,s_0)}{s_0}\right)
= e^{-2(t_1-t_0)}s_1
\left(1-\frac{\vp(t_1,s_1)}{s_1}\right). 
\end{equation}
We take $t_0=T_1$ and apply (\ref{eq:decay_u_add4}) and the hypothesis
$|s_1| \leq Re^{2(t_1-T)}=\frac{1}{2}Re^{2(t_1-T_1)}$ to deduce 
\[ R < |s_0| \leq  |s_1| e^{-2(t_1-T_1)} \frac{1 + 1/3}{1- 1/3} 
 \leq R, \]
a contradiction.  This proves the claim.

We now apply these claims to propagate the decay estimate
(\ref{eq:decay_u_add3}). From claim 2, for any $t=t_1 \geq T$, 
$R \leq k \leq Re^{2(t-T)}$, with $s_1=ik$ we obtain 
$t_0\in[T_1,t]$ and $s_0\in C_R$ and substitute  
(\ref{eq:phi_time2}), (\ref{eq:decay_u_add3}) and (\ref{eq:decay_u_add4}) 
in the decay estimate (\ref{eq:add_str10}) to obtain 
\begin{eqnarray}
\nonumber
|v(t,ik)| &\leq & 
\frac{|v(t_0,s_0)|}{(1-|u(t_0,s_0)|)^{3}}  
\left| \frac{s_0 - \vp(t_0,s_0)}{ik-\vp(t,ik)}\right|^{3/2} 
\\ \nonumber
& \leq &
    (1-\veps)^{-3} \  |v(t_0,s_0)| \left| \frac{ 2s_0}{k} \right|^{3/2} 
\\ \nonumber
\label{eq:v_unif_est}
&  \leq& (1-\veps)^{-3} (2R)^{-3/2} 
  \left( \frac{2R}{k} \right)^{3/2} =(1-\veps)^{-3} k^{-3/2}.
\end{eqnarray}
Therefore, 
\begin{equation}\label{vest2}
\int_R^{Re^{2(t-T)}} |v(t,ik)| \,dk \leq
(1-\veps)^{-3} \int_R^{\infty} k^{-3/2}\, dk =  
\frac{2 R^{-1/2}}{(1-\veps)^{3}}
\leq C\veps,  
\end{equation}
with $C = 2 (8/7)^3 2^{1/2}$. This completes the proof of
Lemma~\ref{le:transition}.

\subsection{Proof of Lemma~\ref{le:contour_decay}}
We consider the initial time $t_0=0$  and the following special case of
(\ref{eq:char_soln}):  
\begin{equation}
\label{eq:flow_phi}
s = s(t;0,s_0) = e^{2t}\left[ s_0 -
  \varphi_0(s_0)(1-e^{-t})\right].
\end{equation}
For any $t \geq 0$, the map $s_0 \mapsto s(t;0,s_0)$ is analytic
for Re($s_0)>0$, and 
\begin{equation}
\label{eq:der_phi}
\frac{ds}{ds_0} = e^{2t} \left(1 - u_0(s_0)(1-e^{-t})
\right), \qquad u_0(s_0)=u(0,s_0).
\end{equation}
Recall that $\dom_t$ denotes the image of $\Cint$
under $s_0 \mapsto s(t;0,s_0)$, and $\Gamma_t$ denotes the image of
the imaginary axis; we let $\Gammamt$ denote its preimage.   As was
observed in Lemma~\ref{le:char_map}, $\Gamma_t$ is a graph over the
imaginary axis, contained in the left half plane.

We will use the
analyticity of $v(t,s)$ in $\dom_t$ and contour deformation.
\begin{figure}
\label{fig:strong1}
\centerline{\epsfysize=7
cm{\epsffile{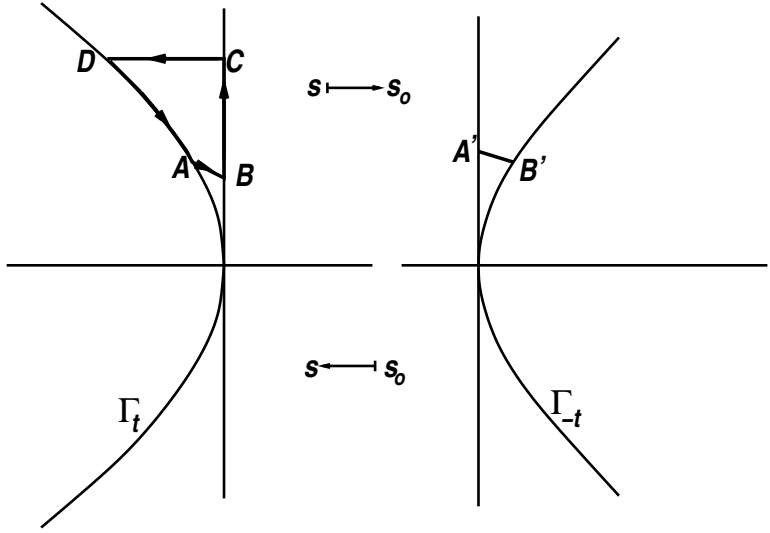}}}
\caption{The $s$-plane is on the left, the $s_0$-plane on the
  right. $\dom_t$ is the region to the right of $\Gamma_t$.
$A= s(t;0,i\tilde{R})$, $B= i\tilde{R}e^{2t}$, $C=iR_2$,
  $\mbox{Im}(D)=R_2$, $A'= i\tilde{R}$, $B'=s(0;t, i\tilde{R}e^{2t})$} 
\end{figure}
For  large finite $R_2 <\infty$ consider the domain $ABCD$ shown in
Figure~\ref{fig:strong1}. The path $AB$ is chosen so that
$A'B'$ is a straight line. $CD$ is parallel to the real
axis, and lies in $\dom_t$ since $\Gamma_t$ is a graph over the imaginary
axis. Then by Cauchy's theorem,
\begin{eqnarray*}
&&  \int_{\tilde{R}e^{2t}}^{R_2}  e^{ikx} v(t,ik)\,dk  =
\int_{BC} e^{ikx}v(t,ik)\,dk
\\ &&\quad = \ 
  \int_{DA} e^{sx} v(t,s)\,ds  + \int_{AB} e^{sx} v(t,s)
  \,ds   + \int_{CD} e^{sx} v(t,s) \,ds . 
\end{eqnarray*}
Let $\sigma$ denote $\Re s$. Since $\sigma <0$ in $\dom_t$ for $s \in
CD$ we see that the last integral is estimated by
\[ \left| \int_{CD} e^{sx} v(t,s) \,ds \right| \leq \sup_{s \in CD}
|v(t,s)| \int_{-\infty}^0 e^{\sigma x} d\sigma = \frac{\sup_{s \in CD}
  |v(t,s)|}{x}. \]
By the decay estimate (\ref{eq:add_str10}) we have
\[ \sup_{s \in CD} |v(t,s)| \leq \sup_{s_1 \in CD}
\frac{|v_0(s_0)|e^{-3t}}{\left( 1  -|u_0(s_0)|\right)^3}, \quad
s_0=s(0;t,s_1). \]
It follows from \qref{udecay} and the fact that
$|s_0|=|s_1|e^{-2t}(1+o(1))\to\infty$ as $R_2 \rightarrow \infty$ 
that $\sup_{s_1 \in CD}  |v_0(s_0)| \rightarrow
0$. We thus let $R_2 \rightarrow \infty$ to conclude that
\begin{equation}
\label{tail1}
 \int_{\tilde{R}e^{2t}}^\infty e^{ikx} v(t,ik) \,dk  =
 \int_{\Gamma_{t,A}} 
e^{sx} v(t,s) \,ds +  \int_{AB} e^{sx} v(t,s) \,ds .
\end{equation}
where $\Gamma_{t,A}$ denotes  the path from $\infty$ to $A$ on $\Gamma_t$.
Notice that (\ref{tail1}) holds independent of $x$.

The virtue of deforming the contour is that the integrals may now be
estimated by changing variables from $s$ to $s_0$. 
We use the solution formula (\ref{eq:v_soln}) together with
the change of variables $s=s(t;0,ik)$ and (\ref{eq:der_phi}) to obtain
\begin{eqnarray}
\nonumber
\int_{\Gamma_{t,A}} e^{sx} v(t,s) \,ds = 
i e^{-t}  \int_{\tilde{R}}^\infty   e^{s(t;0,ik)
  x}\frac{v_0(ik)}{\left(1-u_0(ik)(1-e^{-t}) \right)^2}   \,dk.
\end{eqnarray}
Since $\Re s(t;0,ik)\le 0$ and $\sup_{|k|\ge\tilde{R}}|u_0(ik)|<1$, 
this yields the estimate
\begin{equation}\label{tail2}
\left| \int_{\Gamma_{t,A}} e^{sx} v(t,s) \,ds \right| \leq 
C_1 e^{-t} \|v_0\|_{L^1}.
\end{equation}
Similarly, we have by (\ref{eq:v_soln}) and (\ref{eq:der_phi})
\begin{eqnarray}
\nonumber
\left| \int_{AB} e^{sx} v(t,s) \,ds \right| &=&  e^{-t} \left|
    \int_{A'B'} 
e^{s(t;0,s_0)x} \frac{v_0(s_0)}{\left(1-u_0(s_0)(1-e^{-t}) \right)^2}
  \,ds_0 \right|\\[3pt]
\nonumber
 & \leq & e^{-t} |A'B'| \sup_{s_0 \in A'B'}
\left| 1- u_0(s_0)(1-e^{-t})\right|^{-2}.
\end{eqnarray}
The point $A'= i\tilde{R}$ is independent of $t$. It also follows from
(\ref{eq:flow_phi}) that $B' =
s(0;t,i\tilde{R}e^{2t})$ converges to the point $s_0\in\Cclose$ that solves
$i\tilde{R} =s_0 - \varphi_0(s_0)$. Thus, we have the exponential
decay estimate 
\begin{equation}\label{tail3}
\left|\int_{AB} e^{sx} v(t,s) \,ds\right| \leq C_2e^{-t}.
\end{equation}
The constants $C_i$ in (\ref{tail2}) and (\ref{tail3})
depend only on $\tilde{R}$ and the initial data $u_0$. To be explicit,
we set $\tilde{C} = C_1\|v_0\|_{L^1} + C_2$. This completes the proof.

\subsection{The discrete Smoluchowski equations}
We now use  the proof of Theorem~\ref{thm:strong_add} to
obtain a uniform convergence theorem for the discrete Smoluchowski
equations with additive kernel. The proof is simpler and we do not
need the contour deformation argument.

Let $\nu_t = \sum_{l=1}^\infty n_l(t)
\delta_{hl}(x)$ denote a measure-valued solution to
(\ref{eq:smol1}). We first adapt the rescaling (\ref{eq:add_rsc1}) and
(\ref{eq:add_rsc2}) to similarity variables. Let
\begin{equation}\label{eq:rsc-add}
\hat{l} =  lhe^{-2t},\quad \hn_l(t)= h^{-1}e^{4t}n_l(t)  . 
\end{equation}
Then the discrete Fourier inversion formula analogous to 
(\ref{eq:discrete_inv3}) is
\begin{equation}
\label{eq:add_discrete_inv2}
\hat{l}^2\hn_l(t) 
= \frac{1}{2\pi} \int_{-\pi
e^{2t}/h}^{\pi e^{2t}/h}  e^{i  \hat{l} k} v(t,ik) \,dk. 
\end{equation}
\begin{thm}
\label{thm:add_discrete}
Let $\nu_0 \geq 0$ be a lattice measure with span $h$
such that
$\int_0^\infty x \nu_0(dx) = \int_0^\infty x^2 \nu_0(dx)=1$. Then
with the scaling (\ref{eq:rsc-add}) we have
\[ \lim_{t \rightarrow \infty}\, \sup_{l\in\N}\,
\hat{l}^2 \left| \hn_{l}(t)  -
\nsg1(\hat{l}) \right|=0.\]
\end{thm}
\begin{proof}
By (\ref{eq:add_discrete_inv2}) and the  continuous Fourier inversion
formulas it suffices to show that 
\begin{equation}\label{eq:disc-lim}
\lim_{t \rightarrow \infty} \sup_{\hat{l}\ge0}
\left| \int_{-\pi e^{2t}/h}^{\pi e^{2t}/h}
  e^{i  \hat{l} hk} v(t,ik) \,dk 
  - \int_{\mathbb{R}} e^{i\hat{l}h k} v_{*,1}(ik) \,dk \right| =0.
\end{equation}
Let $\veps \in (0,\frac18)$ and choose $R= \frac{1}{2}\veps^{-2}$. The
  integrals over 
  $[-R,R]$ and $R < |k| <   \tilde{R} e^{2t}$ with $\tilde{R}=e^{-2T}$
  are controlled as 
in the proof of Theorem~\ref{thm:strong_add}, and it only remains to
  control the integral of $|v(t,ik)|$ over 
  $\tilde{R}e^{2t} < k < \pi e^{2t}/h$. This is considerably simpler than
in the previous proof. We use the solution formula (\ref{eq:v_soln}) and 
change variables via $ik=s(t;0,s_0)$, using (\ref{eq:der_phi}) to obtain
\begin{eqnarray}
\nonumber
\int_{\tilde{R}e^{2t}}^{\pi e^{2t}/h} e^{ikx} v(t,ik) \,dk 
& = & 
i e^{-t} \int_{\Gammamt(\tilde{R},\pi/h)}
\frac{e^{xs(t;0,s_0)} v_0(s_0)}{\left( 1-u_0(s_0)(1-e^{-t})
  \right)^2} \,ds_0. 
\end{eqnarray}
Here $\Gammamt(\tilde{R},\pi/h)$ denotes the segment along
the curve $\Gammamt$ from
$s(0;t,i\tilde{R}e^{2t})$ to $s(0;t,i\pi e^{2t}/h)$. 
The formula (\ref{eq:char_soln}) shows that
$\Gammamt(\tilde{R},\pi/h)$ converges to
a compact $C^2$ curve defined implicitly by $ik= s_0 -
\varphi_0(s_0), \tilde{R} \leq k \leq \pi/h$. Thus, for $t \geq T$ we
have  
\[  e^{-t} \left| \int_{\Gammamt(\tilde{R},\pi/h)}
\frac{e^{x s(t;0,s_0)} v_0(s_0)}{\left( 1-u_0(s_0)(1-e^{-t})
  \right)^2} \,ds_0 \right| \leq C(T,\tilde{R},u_0,v_0) e^{-t}. \]
Thus, this term is less than $\veps$ for all $t$ large enough.
\end{proof}

\section{Self-similar gelation for the multiplicative kernel}
For $K=xy$, McLeod solved the coagulation equation explicitly for
monodisperse initial data, and showed that a mass-conserving solution failed to
exist for $t >1$. The second moment satisfies
$m_2(t)=(1-t)^{-1}$. The divergence of the second moment indicates
that breakdown is associated with an explosive flux of mass toward 
large clusters.  A rescaled limit of McLeod's solution is the 
following self-similar solution for $K=xy$~\cite{Aldous}:
\begin{equation}
\label{eq:McL2}
 n(t,x) = \frac{1}{\sqrt{2 \pi}} x^{-5/2} e^{-(1-t)^2 x/2}, 
 \quad x>0 ,\quad t < 1.
\end{equation}
Evidently this solution has infinite mass (first moment). 
This should not be thought unnatural, however, since it was shown in
\cite{MP1} that equation (\ref{eq:smol1}) has a unique weak solution
for any initial distribution with finite second moment.

The problem of solving Smoluchowski's equation 
with multiplicative kernel can be reduced to that for the additive kernel by a
change of variables~\cite{Drake}. Let us briefly review this. In
unscaled variables we define
\begin{equation}
\label{eq:mult_phi}
\etamult(t,z) = \int_0^\infty (1-e^{-zx}) x
 n(t,x) \,dx.
\end{equation}
Then $\etamult$ solves the inviscid Burgers equation:
\begin{equation}
\label{eq:mult3}
\partial_t \etamult -\etamult \partial_z \etamult=0,
\end{equation}
with initial data
\begin{equation}
\etamult_0(z) =\int_0^\infty (1 - e^{-zx}) x n_0(x) \,dx.
\end{equation}
The gelation time for initial data with finite 
second moment is $T_{\rm gel}= (\int_0^\infty x^2 \nu_0(dx))^{-1}$ and
this is exactly the time for the first intersection of
characteristics~\cite{MP1}.  We presume that the initial data is
scaled to ensure
\begin{equation}
\la{eq:moment_mult}
\int_0^\infty x^2 n_0(x) \,dx = \int_0^\infty x^3 n_0(x) \,dx = 1.
\end{equation}
Then the gelation time is $T_{\rm gel}=1$.
The connection between the additive and
multiplicative kernels is that $\etamult$ solves (\ref{eq:mult3}) with
initial data $\etamult_0$, if and only if 
$\etaadd(\tau,z)$ is a
solution to (\ref{eq:phi_burgers}) with the same initial data, where 
\begin{equation}
\label{eq:mult4}
\etamult(t,z) = 
e^{\tau} \etaadd(\tau,z) ,
\quad\mbox{with $\tau = \log(1-t)^{-1}$.} 
\end{equation}
For solutions $n^{\rm mul}(t,x)$ and $n^{\rm add}(\tau,x)$ 
to Smoluchowski's equation
with multiplicative and additive kernels respectively, this means that 
\begin{equation}\label{eq:m=add}
xn^{\rm mul}(t,x)= (1-t)^{-1}n^{\rm add}(\tau,x)
\end{equation}
for all $t\in(0,1)$, if and only if the same holds at $t=0$.
We thus obtain a scaling limit as $t\to\Tgel$ directly from
Theorem~\ref{thm:strong_add}. The similarity variables for the
multiplicative kernel are
\begin{equation}
\label{eq:mult_rescaling3}
\hat{x} = (1-t)^2 x , \quad \hat{n}(t,\hat{x}) = 
\frac{n(t, \hat{x}(1-t)^{-2}) } {(1-t)^{5}}
= 
\frac{n(t,x)}{(1-t)^{5} },
\end{equation}
and the self-similar profile is
\begin{equation}
\label{eq:mult_rescaling4}
\nsg2(\hat{x}) = \frac{1}{\sqrt{2\pi \hat{x}^5}}e^{-\hat{x}/2}. 
\end{equation}

\begin{thm} 
\label{thm:strong_mult}
Suppose $n_0(x) \geq 0$, $\int_0^\infty x^2n_0(x) \,dx = \int_0^\infty
x^3 n_0(x) \,dx = 1$. Suppose also that the Fourier transform of 
 $x^3 n_0$ is integrable. Then in terms of the rescaling
(\ref{eq:mult_rescaling3}) we have
\begin{equation}
\label{eq:strong_fourier_mult}
\lim_{t \to 1} \,\sup_{\hat{x}>0}\,
\hat{x}^3| \hat{n}(t,\hat{x}) - \nsg2(\hat{x}) |  =0,
\end{equation}
where $\nsg2(\hx)$ is the self-similar density in
(\ref{eq:mult_rescaling4}). 
\end{thm}

Notice that (\ref{eq:mult_rescaling3})
is {\em not\/} a mass-preserving rescaling; 
indeed, the rescaled mass diverges:  
\[ \int_0^\infty \hat{x} \hat{n}(t,\hat{x}) d\hat{x} = \frac{1}{1-t}
\int_0^\infty x n(t,x) \,dx =\frac{1}{1-t} \to \infty. \]
Instead, (\ref{eq:mult_rescaling3}) preserves the second moment:
\[ \int_0^\infty \hat{x}^2 \hat{n}(t,\hat{x}) d\hat{x} = (1-t)
\int_0^\infty x^2 n(t,x) \,dx = 1, \quad t \in [0,1).\]
The explanation is that the scaling in
(\ref{eq:mult_rescaling3}) is designed to capture the behavior of
the distribution of large clusters as $t$ approaches $\Tgel$ ---
the average cluster size is $(1-t)^{-1}$.
Correspondingly, the mass of the
self-similar solution is infinite. 

Theorem~\ref{thm:add_discrete} may be similarly adapted to $K=xy$. 
In the discrete case, the correspondence (\ref{eq:m=add}) between
solutions of Smoluchowski's equations with multiplicative and additive
kernels becomes
\begin{equation}\label{eq:m-add-disc}
hl n_l^{\rm mul}(t) = (1-t)^{-1} n^{\rm add}_l(\log(1-t)^{-1})
\end{equation}
We introduce similarity variables via
\begin{equation}\label{eq:rsc-m}
\hat{l} = lh(1-t)^2, \quad \hat{n}_l(t) = h^{-1}(1-t)^{-5}n_l(t) .
\end{equation}
Then directly from Theorem~\ref{thm:add_discrete}
we obtain the following.

\begin{thm}
\label{thm:mult_discrete}
Let $\nu_0 \geq 0$ be a lattice measure with span $h$ such that
$\int_0^\infty x^2 \nu_0(dx) = \int_0^\infty x^3 \nu_0(dx)=1$. Then
with the rescaling (\ref{eq:rsc-m}) we have
\begin{equation}
\label{eq:mult_lim}
\lim_{t \rightarrow 1}\, \sup_{l\in\N}\, \hat{l}^3 \left| 
\hat{n}_{l}(t) - \nsg2(\hat{l}) \right|=0.
\end{equation}
\end{thm}

\section*{Acknowledgements}
The authors are grateful to an anonymous referee for suggestions
that greatly improve the presentation and its accuracy.
The authors thank the Max Planck Institute for Mathematics in the
Sciences, Leipzig for hospitality during part of this work. 
G.M. thanks Timo Sepp\"{a}l\"{a}inen for his help during early stages
of this work.
This material is based upon work supported by the National Science
Foundation under grant nos.\ DMS 00-72609 and DMS 03-05985.

\bibliographystyle{siam}
\bibliography{smol2}
\end{document}